# Generation of Switchable Singular Beams with Dynamic Metasurfaces

*Ping Yu[1,†], Jianxiong Li[1,†], Xin Li[2], Gisela Schütz[1], Michael Hirscher[1,]\*, Shuang Zhang[3,]\*, and Na Liu[1,4]\**

[1]Max Planck Institute for Intelligent Systems, Heisenbergstrasse 3, 70569 Stuttgart, Germany

[2]Beijing Engineering Research Center for Mixed Reality and Advanced Display, School of Optoelectronics, Beijing Institute of Technology, South Zhongguancun Street 5, 100081 Beijing, China

[3]School of Physics & Astronomy, University of Birmingham, Birmingham B15 2TT, UK.

[4]Kirchhoff Institute for Physics, University of Heidelberg, Im Neuenheimer Feld 227, 69120 Heidelberg, Germany.

**ABSTRACT:** Singular beams have attracted great attention due to their optical properties and broad applications from light manipulation to optical communications. However, there has been a lack of practical schemes to achieve switchable singular beams with subwavelength resolution using ultrathin and flat optical devices. In this Letter, we demonstrate the generation of switchable vector and vortex beams utilizing dynamic metasurfaces at visible frequencies. The dynamic functionality of the metasurface pixels is enabled by utilization of magnesium nanorods, which possess plasmonic reconfigurability upon hydrogenation and dehydrogenation. We show that switchable vector beams of different polarization states and switchable vortex beams of different topological charges can be implemented through simple hydrogenation/dehydrogenation of the same metasurfaces. Furthermore, we demonstrate a two cascaded metasurface scheme for holographic pattern switching, taking inspiration from orbital angular momentum-shift keying. Our work provides an additional degree of freedom to develop high-security optical elements for anti-counterfeiting applications.

Vector and vortex beams are the forms of singular beams based on spatial distributions of polarization and phase, respectively. A vector beam possesses polarization singularity, which is characterized by a space-variant polarization distribution in the transverse plane. A vortex beam manifests phase singularity, which is characterized by a helical phase-front, carrying an orbital angular momentum (OAM). Singular beams empower a broad range of applications in particle trapping and manipulation,[1] high-resolution imaging[2] and lithography as well as quantum science and optical communications.[3-7] Optical fibers,[8] *q*-plates,[9] and spatial light modulators[10,11] have been adopted to generate singular beams. Nevertheless, the optical elements involved in these approaches are usually bulky and complex. Such fundamental and technological difficulties pose challenges to the development of compact, efficient, multifunctional optical devices for integrated optics.

Metasurfaces provide an elegant solution to these difficulties. A metasurface is an artificial nanostructured interface that manipulates light by spatially arranged meta-atoms. These meta-atoms, usually consisting of plasmonic or dielectric nanoantennas, can locally control light properties such as phase, amplitude, and polarization.[12,13] Metasurfaces have enabled a family of ultrathin and flat optical elements, which lead to a plethora of optical functionalities including light focusing and steering,[14] holography,[15,16] imaging,[17-19] vector and vortex beam generations,[20-24] among others. However, research endeavors have been mainly devoted to static metasurface devices, in which the properties of individual pixels cannot be reconfigured in real time, especially at visible frequencies. Although in previous studies tunability has been demonstrated by changing the propagation direction or polarization state of the incident light, the metasurfaces themselves are intrinsically static, once the devices are fabricated.[25-27] This leaves out many opportunities that metasurfaces can offer.

In this Letter, we demonstrate the generation of switchable singular beams utilizing dynamic metasurfaces at visible frequencies. The dynamic functionality of the metasurface pixels is enabled by utilization of magnesium (Mg) nanorods, which possess plasmonic reconfigurability upon hydrogenation and dehydrogenation.[28,29] Mg can undergo a phase-transition from metal to dielectric upon $H_2$ loading, forming magnesium hydride ($MgH_2$). This transition is reversible through dehydrogenation using $O_2$. We show that switchable vector beams of different polarization states and switchable vortex beams of different topological charges can be accomplished through simple hydrogenation/dehydrogenation of the metasurfaces without physically changing any optical element. We further demonstrate a proof of concept experiment using dynamic metasurfaces for holographic pattern switching through OAM multiplexing. This work features a paradigm for realization of compact and multi-tasking dynamic optical elements.

**RESULTS AND DISCUSSION:**

We start with the design of a plasmonic metasurface, which can generate two anomalously reflected vortex beams with opposite topological charges of l and –l, respectively, at normal incidence of circularly polarized (CP) light as shown in Figure 1a. Upon incidence of right-handed circularly polarized (RCP) light, the two reflected vortex beams (off-axis angle = 25°) with the same polarization can be written as follows:

$$\frac{1}{2}\exp[i(l\theta + \varphi_1)]\begin{bmatrix}1\\-i\end{bmatrix} \quad (1)$$

$$\frac{1}{2}\exp[i(-l\theta + \varphi_2)]\begin{bmatrix}1\\-i\end{bmatrix} \quad (2)$$

, where $\theta$ is the azimuthal angle. $\varphi_1$ and $\varphi_2$ are the initial phases of the two vortex beams, respectively. $\begin{bmatrix}1\\-i\end{bmatrix}$ is the Jones vector of RCP light. Upon incidence of left-handed circularly polarized (LCP) light, the phase profiles flip in the spatial domain,[15] forming two vortex beams as follows:

$$\frac{1}{2}\exp[i(-l\theta + \varphi_2)]\begin{bmatrix}1\\i\end{bmatrix} \quad (3)$$

$$\frac{1}{2}\exp[i(l\theta + \varphi_1)]\begin{bmatrix}1\\i\end{bmatrix} \quad (4)$$

For linearly polarized (LP) light, which contains RCP and LCP light with equal intensity, the reflected light is a vector beam, resulting from the superposition of two vortex beams of opposite topological charges (see Supporting Information).

The values of $\varphi_1$ and $\varphi_2$ control the polarization distributions of the generated vector beam. For example, if $\varphi_1 = \varphi_2$, the reflected light on one side (Eq. (1) + Eq. (3)) can be written as $\exp(i\varphi_2)\begin{bmatrix}\cos(l\theta)\\\sin(l\theta)\end{bmatrix}$, corresponding to a radially polarized vector beam. On the other hand, if $\varphi_1 = \varphi_2 + \pi$, the reflected light (Eq. (1) + Eq. (3)) can be written as $\exp[i(\varphi_2 + \pi/2)]\begin{bmatrix}-\sin(l\theta)\\\cos(l\theta)\end{bmatrix}$, corresponding to an azimuthally polarized vector beam. As a result, vector beams of different polarization states (radial or azimuthal polarization) can be generated by properly defining the relationship between the two initial phases.

Next, we demonstrate switching between azimuthal and radial polarizations using a dynamic metasurface. As illustrated by the schematic in Figure 1b, azimuthally and radially polarized vector beams of $l = 1$ are designed as initial and final states, respectively. The switching between these two states is enabled by reversible hydrogenation and dehydrogenation of the metasurface. The working principle of the metasurface is presented in Figure 1c. Each super unit cell of the metasurface contains three pixels: one gold (Au) nanorod and two Mg nanorods. All the nanorods have the same dimensions of 200 nm × 80 nm × 50 nm. The Au nanorod remains orthogonal with respect to one of the Mg nanorods (*i.e.*, Mg (I)). Before hydrogenation, these two pixels cancel each other in reflectance in the far field due to destructive interference[30]. The net metasurface functionality is governed by the second Mg nanorod (*i.e.*, Mg (II)), which is geometrically engineered to obtain the acquired Pancharatnam-Berry (PB) phase profiles for generating an azimuthally polarized vector beam of $l = 1$ (see Figure 1b). After hydrogenation, the two Mg nanorods are transformed into $MgH_2$ nanorods. The net function of the super unit cell is thus determined by the Au nanorod, which is engineered to generate a radially polarized vector beam of $l = 1$ (see Figure 1b). As a result, switchable vector beams of different polarization states can be readily achieved through hydrogenation/dehydrogenation of the metasurface without changing any optical elements in the optical path.

Figure 1c shows the scanning electron microscopy (SEM) image of the metasurface sample. The sample fabrication process can be found in Supporting Information. For optical characterizations, the metasurface placed in a gas cell is illuminated by LP light at 633 nm (see the optical setup in Figure S1). The experimental results are presented in Figure 1d. Before hydrogenation (upper row), the intensity profile (first column) shows a typical doughnut shape. When a polarizer is placed in the optical path and rotates continuously, the polarized intensities are petal-shaped, indicating the generation of an azimuthally polarized vector beam of $l = 1$. After hydrogenation the petal-shaped intensity profiles are orthogonal to the corresponding ones in the upper row following the same analyzing polarizer. This confirms the generation of a radially polarized vector beam of $l = 1$. The two polarization states can be reversibly switched using $H_2$ and $O_2$, respectively. It is worth mentioning that to shape the light wavefronts eight nanorods oriented from 0° to 180° at an interval of 22.5° (*i.e.*, eight-phase levels) are utilized to generate the desired phase profile on the metasurface. It results in changes of the light intensity after the rotating analyzer at different polarization angles. This can be improved by increasing the number of phase levels on the metasurface. Videos that record the vector beam switching between azimuthal and radial polarizations with $l = 1, 2$, and 3 can be found in Movie S1-S3 accordingly. The efficiency of the metasurface device reaches 23%, and the hydrogenation/dehydrogenation process takes 3 minutes in total.

To demonstrate the versatility of our approach, we further show switching between vector beams of different orders in Figure 2a. In this case, the Mg nanorod (*i.e.*, Mg (II)) and the Au nanorod are geometrically engineered on the metasurface to generate radially polarized vector beams of $l = 2$ and $l = 1$, respectively. Figure 2b presents the SEM image of the metasurface sample. As shown in Figure 2c, the number of petals in the intensity patterns can be switched between 4 and 2 through hydrogenation and dehydrogenation. A video that records this reversible process can be found in Movie S4.

The same working principal can also be adopted to achieve switchable vortex beams of different topological charges. CP light is chosen as the incident light in order to generate vortex beams (see the optical setup in Figure S2). Vortex beam switching between topological charges of $l = 1$ and 2 as well as $l = 1$ and 3 upon $H_2$ and $O_2$ loading are demonstrated in Figures 3a and b, respectively. It is evident that the central dark area of the light beam becomes larger, when the topological charge increases. As OAM states of different topological charges are mutually orthogonal, they could be used as carriers of different information channels for multiplexing and transmitting multiple data streams, which potentially increase the capacity of free-space optical communication systems.[31-33] For instance, OAM shift keying (OAM-SK) is one of the important schemes in OAM-based optical communications. In this case, the OAM states of vortex beams are regarded as a modulation format, which enables information encryption by dynamic switching of spatial light modulators.[34]

Next, we demonstrate a proof of concept experiment using dynamic metasurfaces for holographic pattern switching as illustrated in Figure 4a. We employ two metasurfaces, M1 and M2, for encoding and decoding the information, respectively. The first metasurface (M1) is designed to generate a vortex beam at incidence of CP light. Upon $H_2$ and $O_2$ loading, the outgoing vortex beam can be dynamically switched between two different OAM states with topological charges of $l = +m$ and $l = -m$, respectively, as shown in Figure 4a. The data carried by the vortex beam are then transmitted in free space, and incident onto the second metasurface (M2). M2 is designed to reconstruct different information in response to the vortex beam of different OAM states. The design details of M2 can be found in Supporting Information.

The experimental setup is depicted in Figure 4b. A dynamic vortex beam is generated by M1 through hydrogenation/dehydrogenation as described above. Then, the anomalously reflected vortex beam passes through M2, generating different target images for the two different vortex beams. As an example, we use M1 to generate a vortex beam, which can be switched between topological charges of $l = +3$ and $l = -3$. M2 is designed to construct OAM multiplexing hologram of letters "X" and "Y", in response to vortex beams of $l = +3$ and $l = -3$ respectively. The simulated and experimental results are presented in Figure 4c. Initially, only "X" is observed, because the output vortex beam from M1 has a topological charge of $l = +3$. After $H_2$ loading, "X" gradually vanishes and "Y" becomes visible, as the topological charge of the output vortex beam is changed to $l = -3$. Upon $O_2$ loading, "X" appears again. A video that records this process can be found in Movie S5. The efficiency of the metasurface device for OAM hologram generation is 5%. It can be largely improved using a reflective metasurface [16] or a dielectric metasurface[35]. It is also noteworthy that the target information in the two channels has been encoded within a single hologram design. This results in information leakage due to the background noise, which lowers the holographic image contrast. Improved algorithms can be used to increase the accuracy of the complex amplitude reconstructions for achieving a better image contrast. Meanwhile, the incomplete destructive interference between the Au and Mg nanorods in the far field leads to undesired intensity residuals, resulting in a low contrast. Optimizations of the Au and Mg nanorod dimensions for better signal cancellations in the far field will largely improve the contrast.

**CONCLUSIONS:**

In conclusion, we have experimentally demonstrated switchable singular beams with dynamic metasurfaces. Vector beams of different polarization states or vortex beams of different topological charges can be dynamically generated simply by hydrogenation and dehydrogenation of the metasurfaces. In particular, we have shown holographic pattern switching using two cascaded metasurfaces, taking inspiration from OAM-SK. For practical applications, the efficiencies of the metasurface devices can be enhanced through further design optimizations[16]. Meanwhile, their stability and durability can be improved through careful material engineering and optimizations using Mg alloys for achieving robust switching[36] as well as using a polytetrafluoroethylene protective layer for avoiding water staining. Our work provides an additional degree of freedom to develop high-security optical elements for anti-counterfeiting applications.

**EXPERIMENTAL METHODS:**

**Sample fabrications**

All the samples in this work were fabricated using multi-step electron-beam lithography (EBL). First, a structural layer composed of Au nanorods and alignment markers were defined in a PMMA resist (Allresist) using EBL (Raith e_line) on a substrate. A 2 nm Cr adhesion layer and a 50 nm Au film were successively deposited on the substrate through thermal evaporation followed by metal lift-off. Next, the substrate was coated with a PMMA resist. Computer-controlled alignment using the Au markers was carried out to define a second structural layer. Subsequently, 3 nm Ti, 50 nm Mg, 5 nm Ti, and 10 nm Pd films were deposited on the substrate through electron-beam evaporation followed by metal lift-off. The samples for vector and vortex beam generations were manufactured on $SiO_2$ (100 nm)/Si substrates for reflection-based metasurfaces. The sample for OAM multiplexing hologram generation was manufactured on a $SiO_2$ substrate for a transmission-based metasurface.

**Optical setups**

The light beam was generated from a laser diode source (633 nm). A linear polarizer (LP) was employed to obtain LP light. A lens and an aperture were utilized to reshape the light beam to a similar size as the sample area in order to avoid undesired reflected light. The light beam was incident on the metasurface sample placed in a homemade gas cell. The reflected vector beams were projected onto the screens in the far field and captured by a camera as shown in Supporting Figure S1. It should be noticed that the generated vector beam on one of the screens was the one with the desired polarization. For the vortex beam generation, an additional quarter wave plate (QWP) was employed to obtain the required incident CP light. According to the metasurface design, only one screen was necessary in this case (see Supporting Figure S2). All the experiments were carried out at 80°C to facilitate the switching upon $H_2$ or $O_2$ loading.

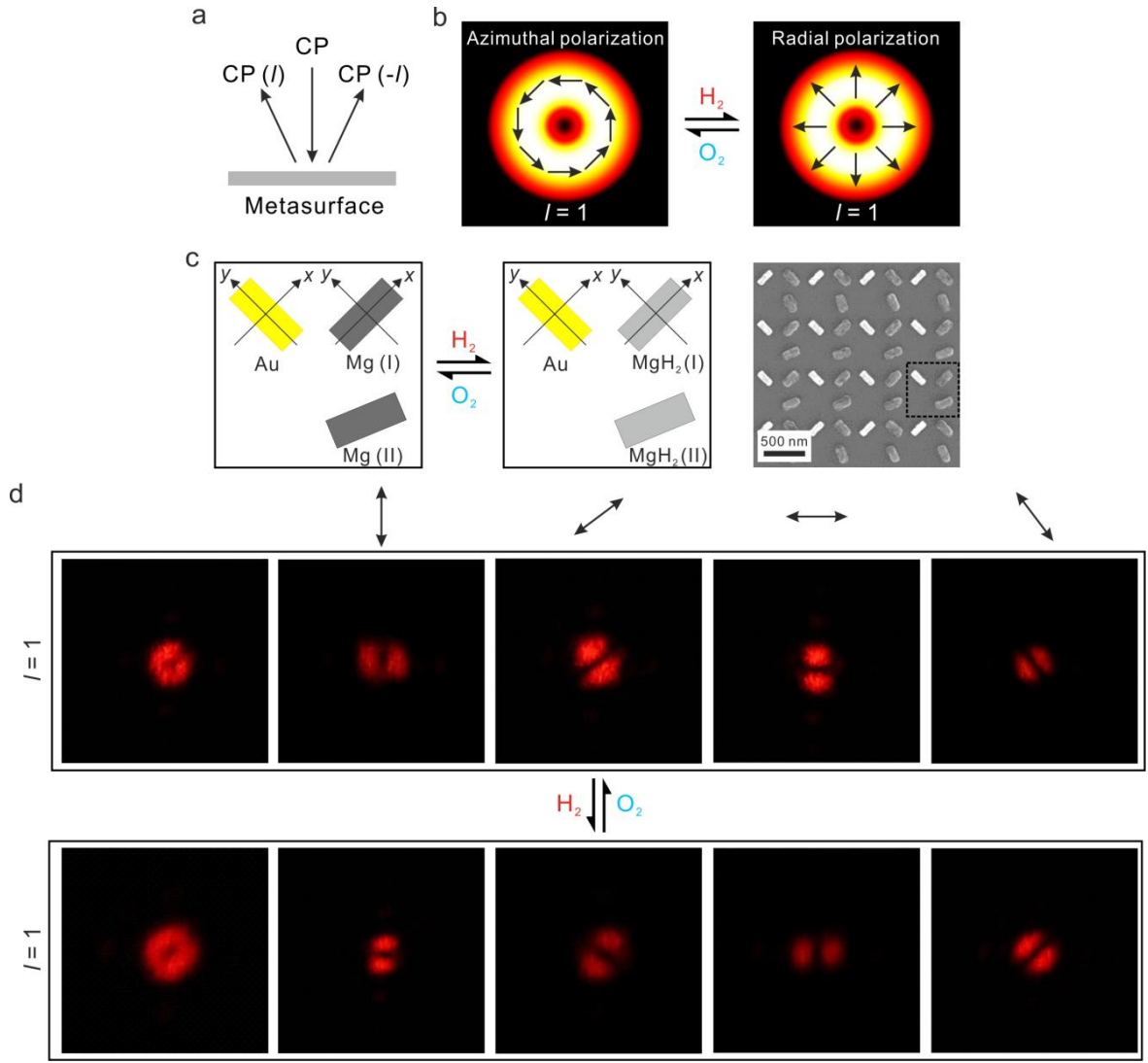

**Figure 1.** Vector beam switching between azimuthal and radial polarizations. (a) Metasurface for generation of vector beams. (b) Schematic illustration of switchable vector beams between azimuthal and radial polarizations of $l$ =1, upon $H_2$ and $O_2$ loading. (c) Working principal and SEM image of the metasurface sample. The dimension of each super unit cell is 600 nm × 600 nm. (d) Experimentally recorded beam patterns, upon switching between azimuthal and radial polarizations through hydrogenation/dehydrogenation. The first column in each row shows a doughnut-shaped intensity profile. The rest of the columns in each row show petal-shaped intensity patterns after passing through a linear polarizer. The orientations of the polarizer are indicated by the arrows above.

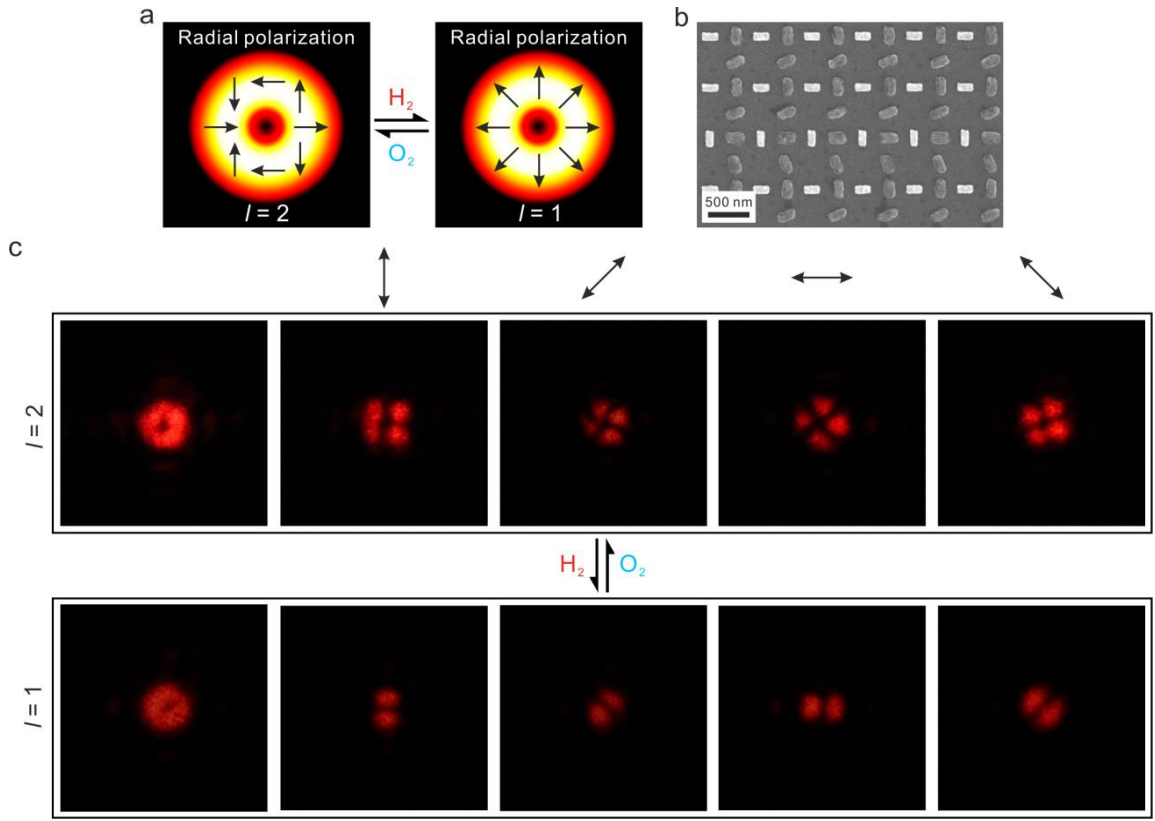

**Figure 2.** Vector beam switching between different polarization orders. (a) Schematic illustration of switchable radially polarized beams between $l = 2$ and $l = 1$, upon $H_2$ and $O_2$ loading. (b) SEM image of the metasurface sample. (c) Experimentally recorded beam patterns, upon switching between $l = 2$ and $l = 1$ through hydrogenation/dehydrogenation. The first column in each row shows a doughnut-shaped intensity profile. The rest of the columns in each row show petal-shaped intensity patterns after passing through a linear polarizer. The orientations of the polarizer are indicated by the arrows above.

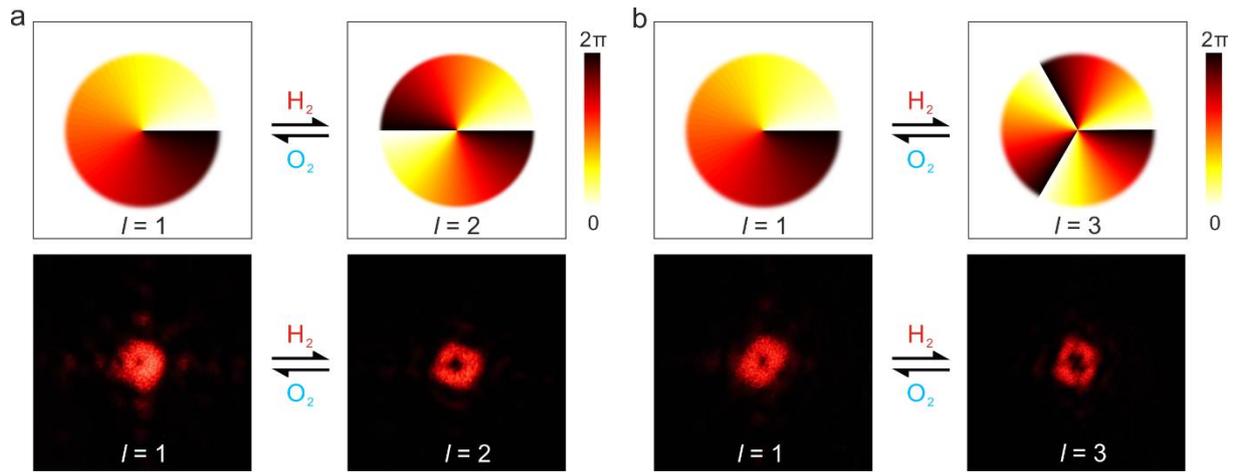

**Figure 3.** Vortex beam switching between different topological charges. (a) Schematic illustration and experimental results of the switchable vortex beams between $l = 1$ and $l = 2$, upon $H_2$ and $O_2$ loading. (b) Schematic illustration and experimental results of the switchable vortex beams between $l = 1$ and $l = 3$ upon $H_2$ and $O_2$ loading.

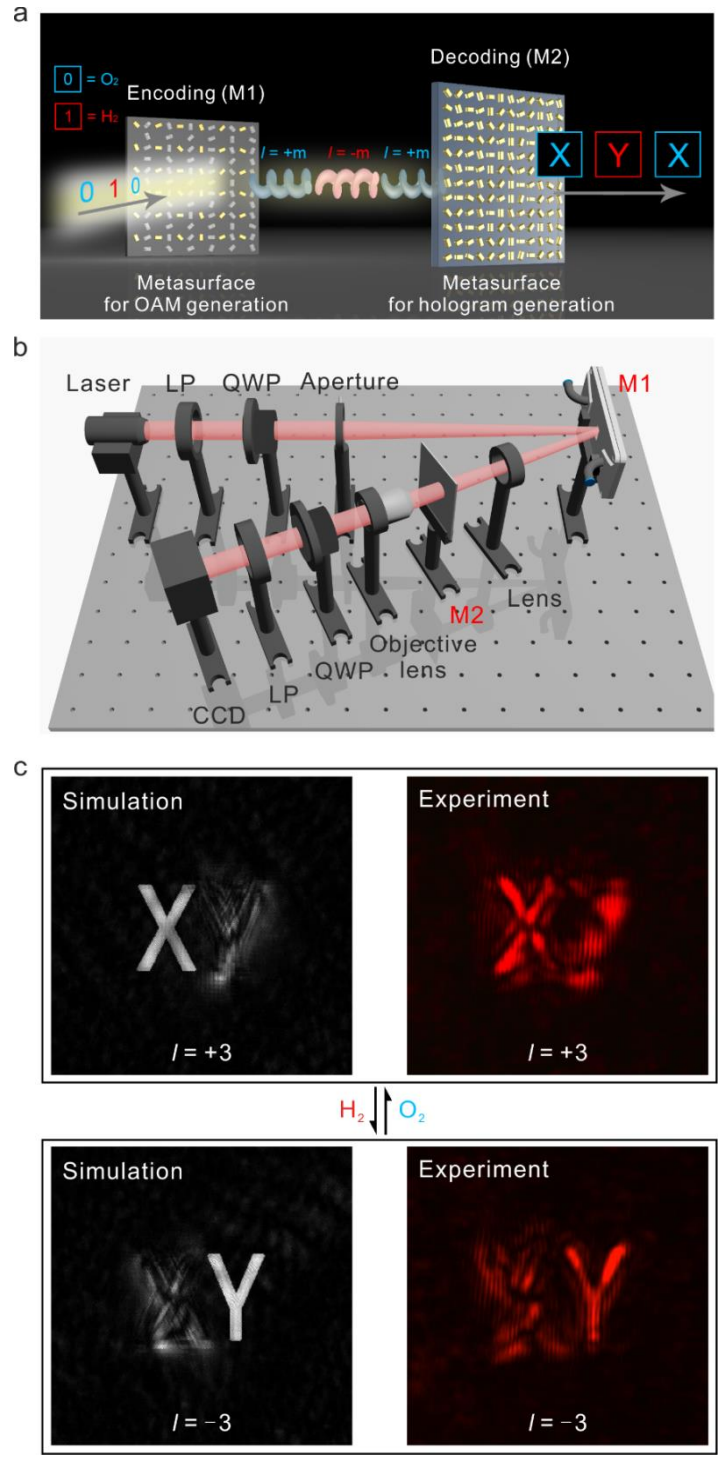

**Figure 4.** Dynamic metasurface for holographic pattern switching through OAM multiplexing. (a) Schematic illustration of the concept. The first metasurface (M1) is designed to generate a vortex beam at incidence of CP light. Upon $H_2$ and $O_2$ loading, the outgoing vortex beam can be dynamically switched between two different OAM states with topological charges of $l = +m$ and $l = -m$, respectively. The second metasurface (M2) is designed to reconstruct different information in response to the vortex beam of different OAM states. (b) Optical setup of the experiment. LP and QWP represent linear polarizer and quarter wave plate, respectively. An objective lens with × 20 magnification is utilized to magnify the images. (c) Simulated and experimental results. Upon $H_2$ and $O_2$ loading, the hologram information, letters "Y" and "X" can be decoded, respectively.

ASSOCIATED CONTENT

**Supporting Information**.

The following files are available free of charge.

Supporting information about the derivations for the vector beam designs and orbital angular momentum (OAM) multiplexing holography designs (word)

Movie S1. Video that records vector beam switching between azimuthal and radial polarizations of $l = 1$. (AVI)

Movie S2. Video that records vector beam switching between azimuthal and radial polarizations of $l = 2$. (AVI)

Movie S3. Video that records vector beam switching between azimuthal and radial polarizations of $l = 3$. (AVI)

Movie S4. Video that records radially polarized vector beam switching between different polarization orders of $l = 2$ and $l = 1$. (AVI)

Movie S5. Video that records the process of holographic pattern switching. (AVI)

AUTHOR INFORMATION

**Corresponding Author**

*Email: hirscher@is.mpg.de

*Email: s.zhang@bham.ac.uk

*Email: na.liu@kip.uni-heidelberg.de

**Author Contributions**

[†] Both authors contributed equally to this work. The manuscript was written through contributions of all authors. All authors have given approval to the final version of the manuscript.

ACKNOWLEDGMENT

This project was supported by the Sofja Kovalevskaja grant from the Alexander von Humboldt-Foundation, the Grassroots Grant from the Max Planck Institute for Intelligent Systems, and the European Research Council (ERC Dynamic Nano and ERC Topological) grants. We gratefully acknowledge the generous support by the Max-Planck Institute for Solid State Research for the usage of clean room facilities.We gratefully acknowledge the generous support by the Max-Planck Institute for Solid State Research for the usage of clean room facilities.

ToC Figure:

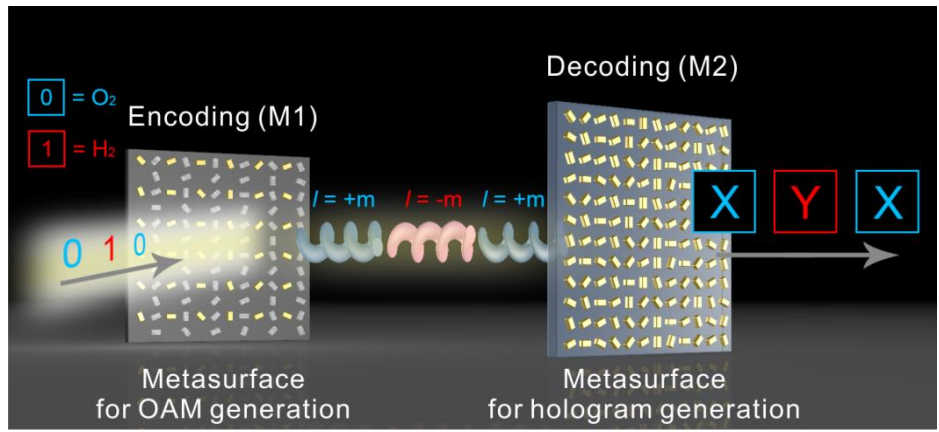